\documentclass[useAMS,usenatbib]{mn2e}
\usepackage{graphicx}

\title[Discovery of a T dwarf + white dwarf binary] 
{Discovery of a T dwarf + white dwarf binary system}
\author[A. C. Day-Jones et al.]{ 
A. C. Day-Jones$^{1,2}$\thanks{E-mail:adjones@das.uchile.cl. Based on observations made with ESO telescopes at the La Silla Paranal Observatory under programme 282.C-5069(A)}, D. J. Pinfield$^2$, M.T. Ruiz$^1$, H. Beaumont$^2$, B. Burningham$^2$, \\ \newauthor J. Gallardo$^1$, A. Gianninas$^3$, P. Bergeron$^3$, R. Napiwotzki$^2$, J.S.Jenkins$^1$, Z.H. Zhang$^2$, \\ \newauthor D. Murray$^2$, S. Catal\'an$^2$, J. Gomes$^2$.\\ $^{1}$ Departamento de Astronomia, Universidad de Chile, Camino del Observatorio 1515, Santiago, Chile. \\ $^{2}$  Centre for Astrophysics Research, University of Hertfordshire, College Lane, Hatfield, Hertfordshire, UK.\\ $^{3}$ D\'epartement de Physique, Universit\'e de Montr\'eal, C.P. 6128, Succursale Centre-Ville, Montreal, Canada.}

\begin{document}
\date{}
\pagerange{\pageref{firstpage}--\pageref{lastpage}} \pubyear{2007}
\maketitle
\label{firstpage}

\begin{abstract}

We present the discovery of the first T dwarf + white dwarf binary system LSPM\,1459+0857\,AB, confirmed through common proper motion and spectroscopy. The white dwarf is a high proper motion object from the LSPM catalogue that we confirm spectroscopically to be a relatively cool ($T_{\rm eff}$ = 5535$\pm$45K) and magnetic (B$\sim$2MG) hydrogen-rich  white dwarf, with an age of at least 4.8\,Gyrs. The T dwarf is a recent discovery from the UKIRT Infrared Deep Sky Survey (ULAS\,1459+0857), and has a spectral type of T4.5$\pm$0.5 and a distance in the range 43-69pc. With an age constraint (inferred from the white dwarf) of $>$4.8\,Gyrs we estimate $T_{\rm eff}$ = 1200-1500K and $\log g$ =5.4-5.5 for ULAS\,1459+0857, making it a benchmark T dwarf with well constrained surface gravity. We also compare the T dwarf spectra with the latest LYON group atmospheric model predictions, which despite some shortcomings are in general agreement with the observed properties of ULAS\,1459+0857. The separation of the binary components (16,500-26,500\,AU, or 365 arcseconds on the sky) is consistent with an evolved version of the more common brown dwarf + main-sequence binary systems now known, and although the system has a wide separation, it is shown to be statistically robust as a non spurious association. The observed colours of the T dwarf show that it is relatively bright in the $z-$ band compared to other T dwarfs of similar type, and further investigation is warranted to explore the possibility that this could be a more generic indicator of older T dwarfs.  Future observations of this binary system will provide even stronger constraints on the T dwarf properties, and additional systems will combine to give a more comprehensively robust test of the model atmospheres in this temperature regime. 

\end{abstract}


\section{Introduction}
\label{intro}

Large scale Near Infrared (NIR) and optical surveys such as the 2-Micron 
All Sky Survey (2MASS), the Sloan Digital Sky Survey (SDSS), and the UKIRT 
Infrared Deep Sky Survey (UKIDSS) are aiding the identification of a rapidly 
increasing number of $'$field$'$ brown dwarfs (BDs) (e.g. \citealt{lodieu07}, 
\citealt{pinfield08}, \citealt{burningham09}), as well as probing down into 
new cooler temperature regimes (\citealt{warren07}, \citealt{burningham08}, 
\citealt{delorme08}, \citealt{burningham09}, \citealt{leggett09}). 
In general the estimation of properties of these BDs (e.g. age, mass, 
metallicity) currently relies on model fitting. However, the models are 
very sensitive to a variety of poorly understood processes in BD atmospheres, 
such as the formation of dust condensates \citep{allard01} and non-equilibrium 
chemistry \citep{saumon07}, and the spectroscopic fitting of atmospheric 
properties ($T_{\rm eff}$, $\log g$, [M/H]) is a major challenge. Crucially, 
the nature of BD evolution means that the mass-luminosity relation depends 
strongly on age, and in the absence of well constrained atmospheric properties 
there is no way to accurately determine mass and age.

Identifying objects where one can pin down these properties independently 
can help aid the calibration of models. BDs that are members of binaries, 
where the primary member has age constraints are good sources of benchmark 
BDs (e.g. \citealt{day-jones08}, \citealt{burningham09}, \citealt{zhang10},
\citealt{faherty09}). In particular white dwarf primaries can provide a hard
lower limit on the age of the system (from the white dwarf cooling age) 
and in the case of  high mass white dwarfs (where the main sequence 
progenitor star will have a short lifetime and the age will thus 
be essentially the same as the cooling age of the white dwarf), could 
provide ages constrained at the 10\% level (\citealt{pinfield06} and 
reference therein).

Binary systems containing a white dwarf and a BD however are
observationally rare, and only a handful of such binaries have been
identified. To date there are only five L dwarf + white dwarf systems
that have been spectroscopically confirmed, GD 165B (L4;
\citealt{zuckerman92}), GD 1400 (L6/7; \citealt{farihi04},
\citealt{dobbie05}), WD\,$0137-349$ (L8; \citealt{maxted06},
\citealt{burleigh06}), PG1234+482 (L0; \citealt{steele07},
\citealt{mullally07}) and PHL 5038B (L8; \citealt{steele09})
previously the latest type BD companion to a white dwarf. There have
also been several BD companions to white dwarfs found as cataclysmic
variable systems (CVs, e.g. \citealt{littlefair08})
, although the nature of such objects means that they my be
less useful in the context of studies of typical BD atmospheres.

We present here results from our ongoing search to identify widely
separated BD companions to white dwarfs, expanding on our earlier
searches of 2MASS and SuperCOSMOS (\citealt{day-jones08}), to include
the first results from our combined search of UKIDSS and SuperCOSMOS. 
We present here the discovery of the first T dwarf + white dwarf binary 
system, which we confirm through common proper motion and spectroscopy.

In Section \ref{ukidss} we describe the ongoing search to identify and
spectroscopically confirm late T dwarfs in the UKIDSS Large Area Survey 
(LAS). In section \ref{pms} we describe our proper motion measurements 
of these confirmed T dwarfs and our techniques to search for binary companions 
to these objects. In section \ref{wdspec} we describe the spectroscopy 
of a white dwarf candidate companion and the resulting constraints on 
its properties. In section \ref{stats} we statistically assess the 
likelihood that our new T dwarf + white dwarf binary system is spurious. 
Section \ref{tproperties} determines constraints for the atmospheric 
properties of the T dwarf, taking advantage of its benchmark age constraints 
(from the white dwarf primary). We also perform some basic spectral model fits to 
the T dwarf spectrum, and compare the resulting predictions. Finally in 
Section \ref{concs} we conclude with further discussion of the system 
as a useful benchmark and comment on the direction of future work.

\section{Finding T dwarfs in the UKIDSS Large Area Survey}
\label{ukidss}

The UKIDSS LAS has been searched for T dwarfs using selection techniques 
based on the observed UKIDSS+SDSS colours of previously identified T dwarfs, 
as well as theoretical predictions for the cooler $T_{\rm eff}$= 400-700K 
regime. The selection techniques used to identify these T dwarfs are 
described in detail in \citet{pinfield08}. In this paper we consider 
T dwarfs that were spectroscopically confirmed (Burningham private comm.) 
by the Summer of 2008 (see \citealt{pinfield08} and a sub-sample of \citealt{burningham10}), and the sky coverage appropriate to this sample includes 
the full LAS second data release, 72\% of the new sky in the third data 
release, and 66\% of the new sky in the fourth data release. In total, this 
sample was identified in 890 square degrees of LAS sky.

\subsection{Follow-up photometry and spectroscopy}
\label{followup}

Candidates are followed up in general with additional imaging in the
near infrared and/or optical.  This allows confirmation of the
expected T-like colours and rules out various forms of contamination
(e.g. faint M dwarfs with low signal-to-noise and blue-ward scattered 
near-infrared colours due to large photometric uncertainty, as well as solar
system objects that can appear as non detections due to their motion). 
This followup has been performed using a variety of telescope/instruments, 
including the Wide Field Camera (WFCAM) and Fast Track Imager (UFTI) 
on UKIRT, and the Long-slit Intermediate Resolution Infrared Spectrograph 
(LIRIS) on the William Herschel Telescope (WHT) (all near infrared), as 
well as the ESO Multi-Mode Instrument (EMMI) and the ESO Faint Object 
Spectrograph and Camera (EFOSC2) on the New Technology Telescope (optical).
Spectroscopic confirmation of the LAS candidates was also achieved
using a number of facilities, including the Near Infrared Camera and
Spectrograph (NIRI) and the Gemini Near Infrared Spectrograph (GNIRS)
on the Gemini telescopes, and the Infrared Camera Spectrograph (IRCS) 
on Subaru. The Near Infrared Camera Spectrograph (NICS) on the Telescopio
Nazionale Galileo (TNG) and the UKIRT imager Spectrograph (UIST) were
also used for the brighter T dwarfs. Details of the follow-up imaging
and spectroscopic strategies, as well as the relevant reduction and
calibration techniques used are further discussed in \citet{lodieu07}, \citet{pinfield08} and \citet{burningham10}. 
Corresponding spectral types were derived using the unified T dwarf 
classification scheme of \citet{burgasser06} with an extension from 
\citet{burningham08} for the latest types.

\subsection{T dwarf distances}
\label{distances}

\begin{figure}
\vspace{4.5cm}
\hspace{2.0cm}
\includegraphics{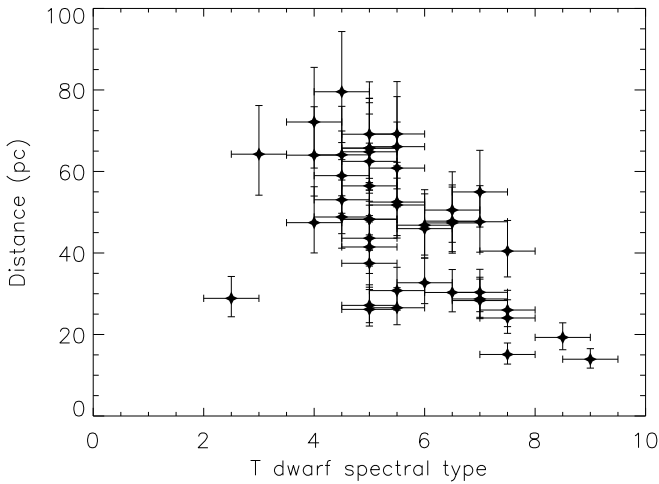}
\vspace{2.0cm}
\caption{T dwarf distance vs spectral type for our spectroscopically confirmed T dwarfs with second epoch imaging.}
\label{tdspt}
\end{figure}

T dwarf distances have been estimated using the absolute
magnitude-spectral type relations from \citet{liu06}, assuming the T
dwarfs are single objects.  We calculated M$_{J}$, choosing the $J-$
band magnitude over the $H-$ and $K-$ bands, as models have suggested
that the $J-$ band may be less sensitive to variations in metallicity
and gravity than the $H-$ and $K-$ bands (e.g.  \citealt{marley02}). The
uncertainties in the distance were obtained by taking into account the
error in the spectral type (typically $\pm$0.5) and the residuals of
the polynomial fits from \citet{liu06}. Fig.~\ref{tdspt} shows the
spectral type distance distribution for the spectroscopically
confirmed sample of 49 T dwarfs that we consider in this work. It can
be seen that the sample spans the spectral type and distance range
T2-9 and 12-80pc. This is comprised of two T2-3 dwarfs, 29 in the T4-6
range, sixteen T6-8 dwarfs and two T8+ dwarfs. Due to the cooler
temperatures and thus fainter nature of later T dwarfs it can be seen
that we are more sensitive to earlier T dwarfs out to further
distances.

\section{T dwarf proper motions}
\label{pms}
\begin{figure}
\vspace{4.5cm}
\hspace{2.0cm}
\includegraphics{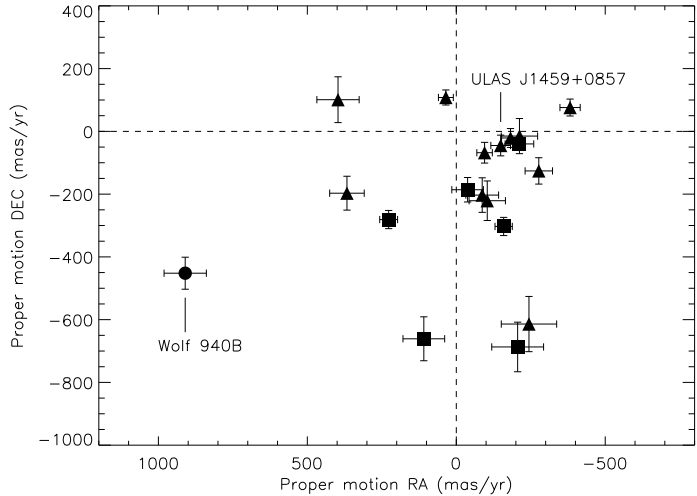}
\vspace{2.0cm}
\caption{A vector point diagram showing the proper motions of our spectroscopically confirmed T dwarfs with good proper motion estimates (see section \ref{pms}). Shown are the different spectral types T4-6, T7-8 and T8+ as squares, triangles and circles, respectively. ULAS\,J1459+0857 and Wolf\,940B are also highlighted.}
\label{tpms}
\end{figure}

The photometric follow-up program provided second epoch imaging for the LAS T dwarfs, which we combined with the LAS images to give two epochs to calculate proper motion. We used the {\sc iraf} routines {\sc geomap} to derive a geometric transformation between the two epoch images, and {\sc geoxytran} to apply these transforms to the T dwarf positions. Centroiding uncertainties were calculated based on simulated data with appropriate Poisson noise injected. The availability and quality of the measured proper motions of the T dwarfs is dependent on several factors including: the baseline between the epochs, the number of stars that can be used for positional reference in each of the images, the S/N of both the T dwarf and the reference stars, and the proximity of the T dwarfs to the edge of the WFCAM detector array (in the first epoch images). We were able to measure the proper motions of 19 T dwarfs from the 49 strong sample of spectroscopic confirmations that we consider, using an average of 12 reference stars across a baseline of 0.5-1.5 years. A vector-point diagram of the T dwarf proper motions is shown in Fig~\ref{tpms}, where two T dwarfs found to have common proper motion companions (see Section \ref{cpm}) are highlighted.

\subsection{A search for common proper motion companions}
\label{cpm}
We searched for candidate common proper motion companions to our
sample of T dwarfs with reliable proper motions, by selecting a magnitude limited sample of SuperCOSMOS sources, where R$<$21 (the magnitude limit at which proper motions are measured) and have
accurately measured proper motions such that PM/$\sigma_{PM}\ge$3. We
searched around each of the 19 T dwarfs out to an angular separation
corresponding to 20,000\,AU, at the estimated minimum distance of each
T dwarf. We choose a separation limit of 20,000AU in order to be sensitive to the detection of both widely separated main-sequence (MS) and white dwarf (WD) companions. It is fairly common to find BD+main-sequence star binaries with separations of $\sim$5000AU (\citealt{gizis01}; \citealt{pinfield06}). However a white dwarf companion could have even wider separations, when one considers any outward migration that would have occurred during the post-main-sequence mass-loss phase. We suggest that the outward migration would likely be up to a factor of $\sim$4, as the initial and final separation of a low mass binary companion is directly related to the change in mass of the host star i.e. $M_{MS}$/$M_{WD}$ (\citealt{jeans1924}; \citealt{zuckerman87}). To illustrate this we consider a white dwarf of mass $\sim$0.65\,M$_{\odot}$ (roughly the mean of the white dwarf mass distribution). The progenitor mass would be $\sim$2.7\,M$_{\odot}$ (from the initial-final mass relations of \citealt{dobbie06}; \citealt{catalan08}; \citealt{kalirai08}), such that M$_{MS}$/M$_{WD}$ $\sim$ 4. Thus, for BD + main-sequence binaries separated by $\sim$5000AU, we could expect their final separation to be up to $\sim$ 20,000AU.

\begin{figure*}
\includegraphics[width=100mm, angle=90]{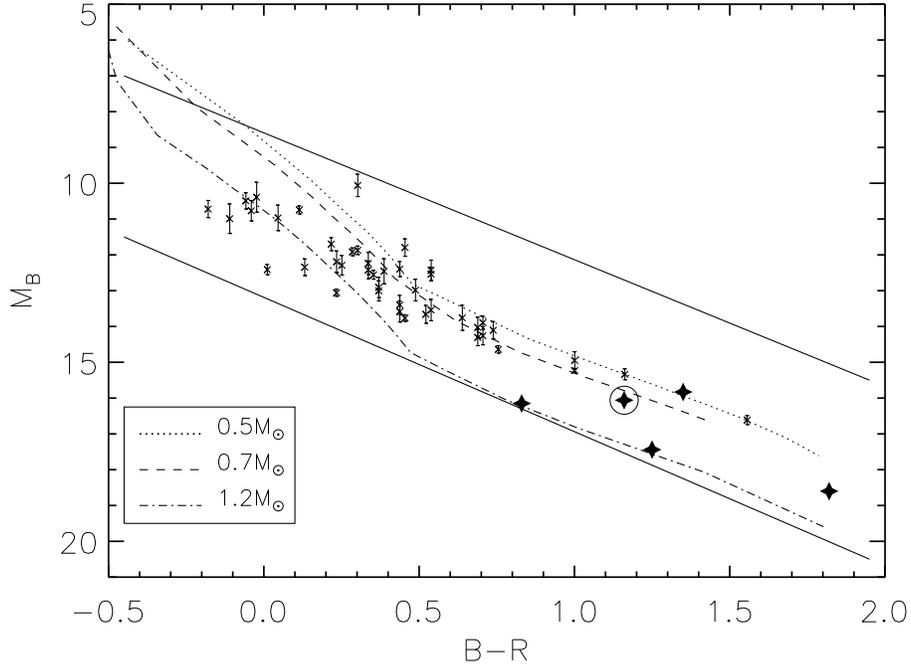}
\caption{A colour-magnitude diagram of white dwarfs from \citet{ms99}
  with known parallax (crosses).  Photometry is on the SuperCOSMOS
  system. Overplotted are model cooling tracks (see main text) for white dwarf
  masses of 0.5, 0.7 and 1.2\,M$_{\odot}$ (dotted, dashed and
  dot-dashed lines respectively). Also overplotted is our white dwarf
  selection region (two solid lines), along with our candidate white
  dwarf companions (large stars). LSPM J1459+0851 is circled for reference.}
\label{wdcmd}
\end{figure*}

\begin{figure*}
\includegraphics[width=100mm, angle=90]{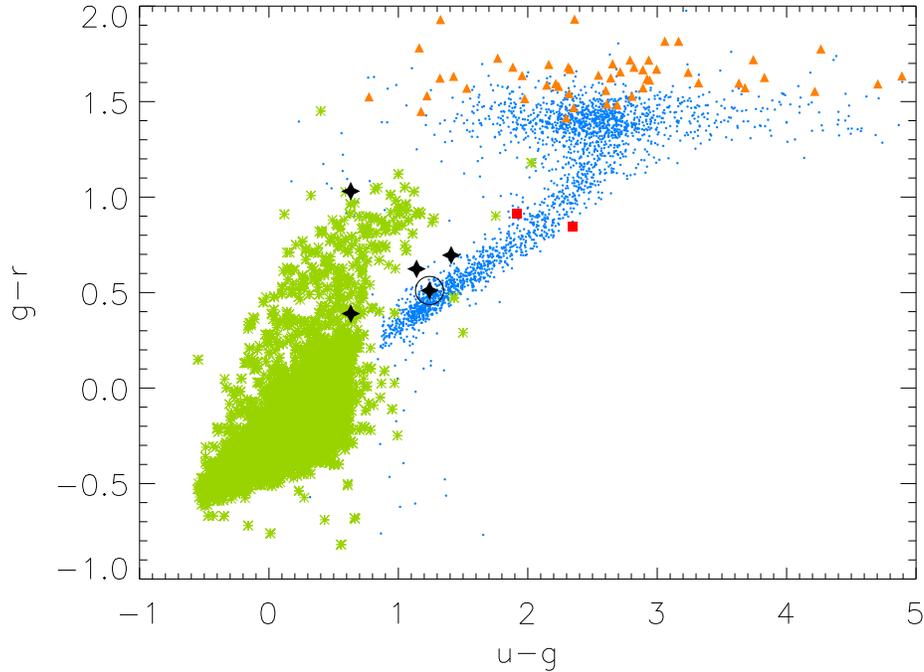}
\caption{A two colour diagram in SDSS colours showing populations
  of main-sequence stars (blue points), M dwarfs (orange triangles),
  white dwarfs (green asterisks) and K subgiants (red squares), with
  our white dwarf candidates overplotted as black stars. LSPM J1459+0851 is circled for reference.}
\label{wd2col}
\end{figure*}

For each common proper motion companion candidate, we assumed a
distance that was the same as the estimated distance of the relevant T
dwarf, and thus placed companion candidates on an absolute B magnitude (M$_{B}$) vs $B-R$
colour-magnitude diagram. Their positions were compared with those
expected for main-sequence stars and white dwarfs, following the
approaches described in \citet{clarke09} and \citet{day-jones08},
respectively.  We then compared their SDSS colours with respect to stellar populations, including main-sequence stars (Hipparcos; \citealt{perryman97}), M dwarfs \citep{west04}, K subdwarfs \citep{yong03} and white dwarfs (\citealt{ms99}, \citealt{eisenstein06}). We found that only one main-sequence companion candidate was identified, Wolf\,940 which had previously been identified serendipitously by \citet{burningham09} and will not be discussed further. We also identified five candidate white dwarf companions to the T dwarfs in our sample, that are common proper motion (to within the uncertainties) and are consistent with the white dwarf sequence (see Fig.~\ref{wdcmd} and \ref{wd2col}) if assumed to be at the same distance as their T dwarf companions.

The brightest of our white dwarf candidates appears in the LSPM
catalogue (LSPM J1459+0851, \citealt{lepine05}) as a high proper
motion object, although it has not been previously studied
spectroscopically. The T dwarf associated with this object is
ULAS\,J1459+0857, which has been spectroscopically typed as a
T4.5$\pm$0.5 dwarf \citep{burningham10}. This pair is highlighted in Fig.~\ref{wdcmd} and \ref{wd2col}.

\subsection{Possible contamination}
\label{contam}
To assess the possibility that our candidate white dwarf companions may be
dominated by high velocity background objects, such as  metal-poor, halo K-subdwarfs (which could populate the same colour and proper motion space that we search in this work), we have derived space motion estimates assuming that our candidates are K subdwarfs. We used the relations of
\citet {ivezic08} to estimate an absolute r$'$ magnitude (M$_{r'}$) from g$'$-i$'$ colour, for a
metallicity range (for sub-dwarfs) of $-$0.5 and $-$2.5, and thus
obtained distance constraints applicable if these objects are
sub-dwarfs. Assuming for simplicity that they have a zero radial
velocity, we then estimated space motions to assess potential halo
membership.  All except one candidate would have space motions of
2,000-12,000~kms$^{-1}$, which are thus not consistent with a galactic halo
population. The exception is close enough such that its kinematics are consistent with a background sub-dwarf member. Indeed this is one of the widest separated candidates (400
arcsec), and the level of background contamination for such separation
(and volume) approaches 1, even for the low density halo luminosity
function \citep{gould03}. However, despite one possible halo contaminant,
the majority of the candidates cannot be explained by such
contamination.

A more likely source of error potentially leading to misidentification amongst the candidates is proper
motion uncertainty, since the ratio of proper motion to proper motion
uncertainty for some of the candidates is in the 3-5 range, and one
would thus expect that some fraction of the sample have proper motions
that have scattered to larger values. Given this likely source of
contamination we choose not to present details of all five of our companion candidates at
this stage. We prefer to first establish their nature through
spectroscopic study, and in that way confirm them (or otherwise) as
genuine white dwarf companions. We have so far only obtained good spectroscopy for both components of the binary containing J1459+0851 and thus focus the remainder of this paper on this one system.

\section{Spectroscopic observations of LSPM J$1459+0851$}
\label{wdspec}

Spectroscopic observations of LSPM J$1459+0851$ were obtained with FORS2 on the Very Large Telescope on 2009 May 15 and 21, with Directors Discretionary Time in program 282.C-5069(A). We used the long slit mode in the optical wavelength range 3300-8000\AA\ with a dispersion of 50 and 55\AA/mm respectively, for the ranges 3300-6210\AA\ (corresponding to the B grism) and 5120-8450\AA\ (corresponding to the RI grism), giving a resolution of R$\sim$1200. Three integrations of 600s were taken, giving a total exposure time of 30 ~minutes for the B grism and two integrations of 360s, totaling 12~minutes in the RI grism. Sky flats, arc frames and the spectra of a DC white dwarf as well as a standard F-type star and DA white dwarf were taken during the same night at a similar airmass to the target so as to provide wavelength, flux and telluric calibrations.

\begin{figure}
\vspace{4.5cm}
\hspace{2.0cm}
\includegraphics{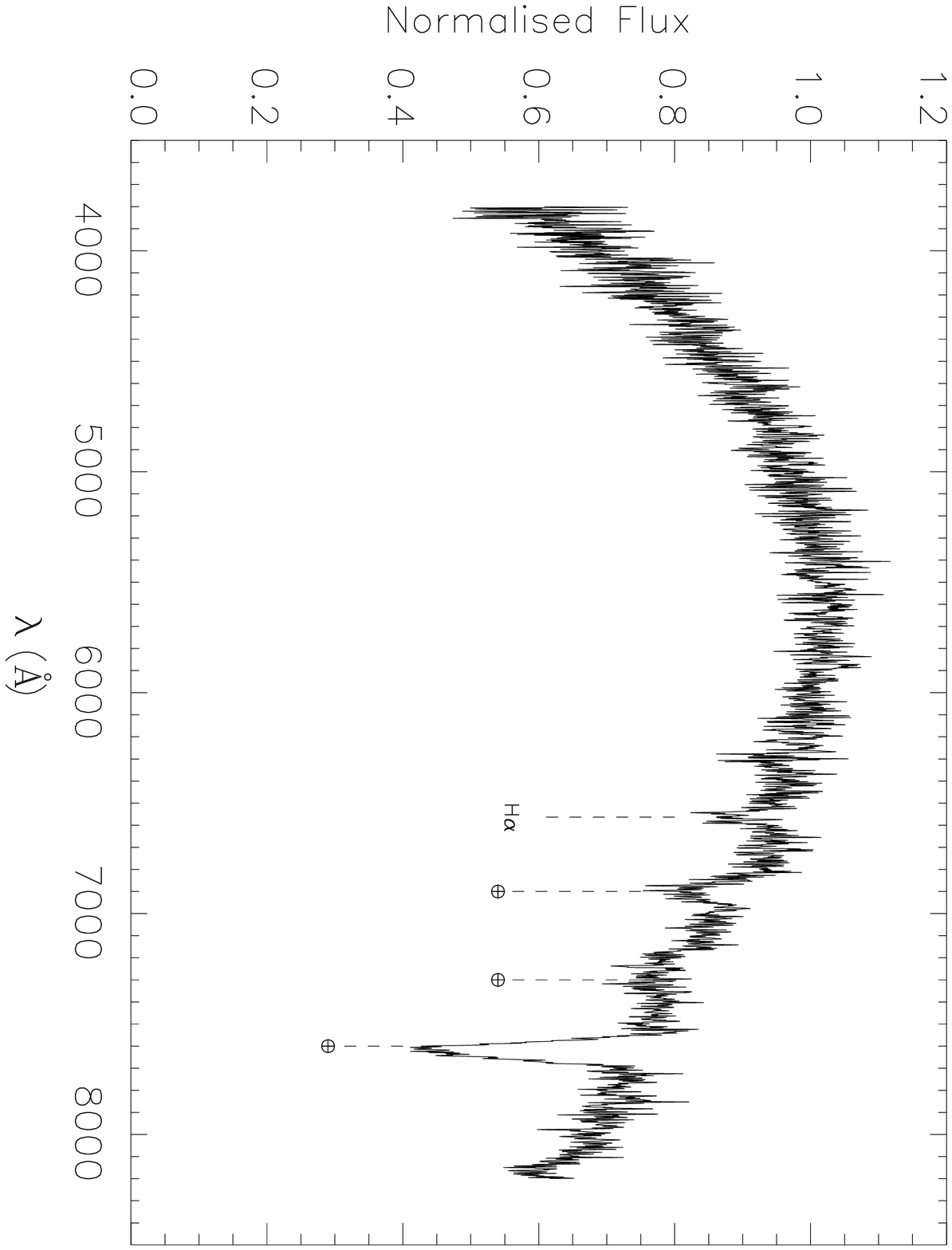}
\vspace{2.0cm}
\caption{The optical spectra of LSPM J$1459+0851$, normalised at 6000\AA. }
\label{wdspecplot}
\end{figure}

Standard {\sc iraf} routines were used to reduce the spectra including flat fielding and cosmic ray removal. The spectra were extracted with {\sc apall}, using a chebyshev function to fit the background and a third order legendre function to trace the fit to the spectrum. The wavelength calibration was done using the spectrum from HgCdHeAr and HgCdHeNeAr arc lamps for the B and RI grisms respectively, and using {\sc identify} to reference the arc lines, along with the {\sc dispcor} routine to correct the dispersion of the spectrum. The resulting spectra of both LSPM J$1459+0851$ and the standard were divided by the smooth spectrum of the DC white dwarf, which has no intrinsic spectral features, enabling correction for the instrumental response. The standard stars (one for each grism) were then used to flux calibrate the spectrum. The two spectra were then stitched together in the overlapping sections and normalised at 6000\AA. The final spectrum of  LSPM J$1459+0851$ is shown in Fig.~\ref{wdspecplot}.

\begin{figure}
\vspace{4.5cm}
\hspace{2.0cm}
\includegraphics{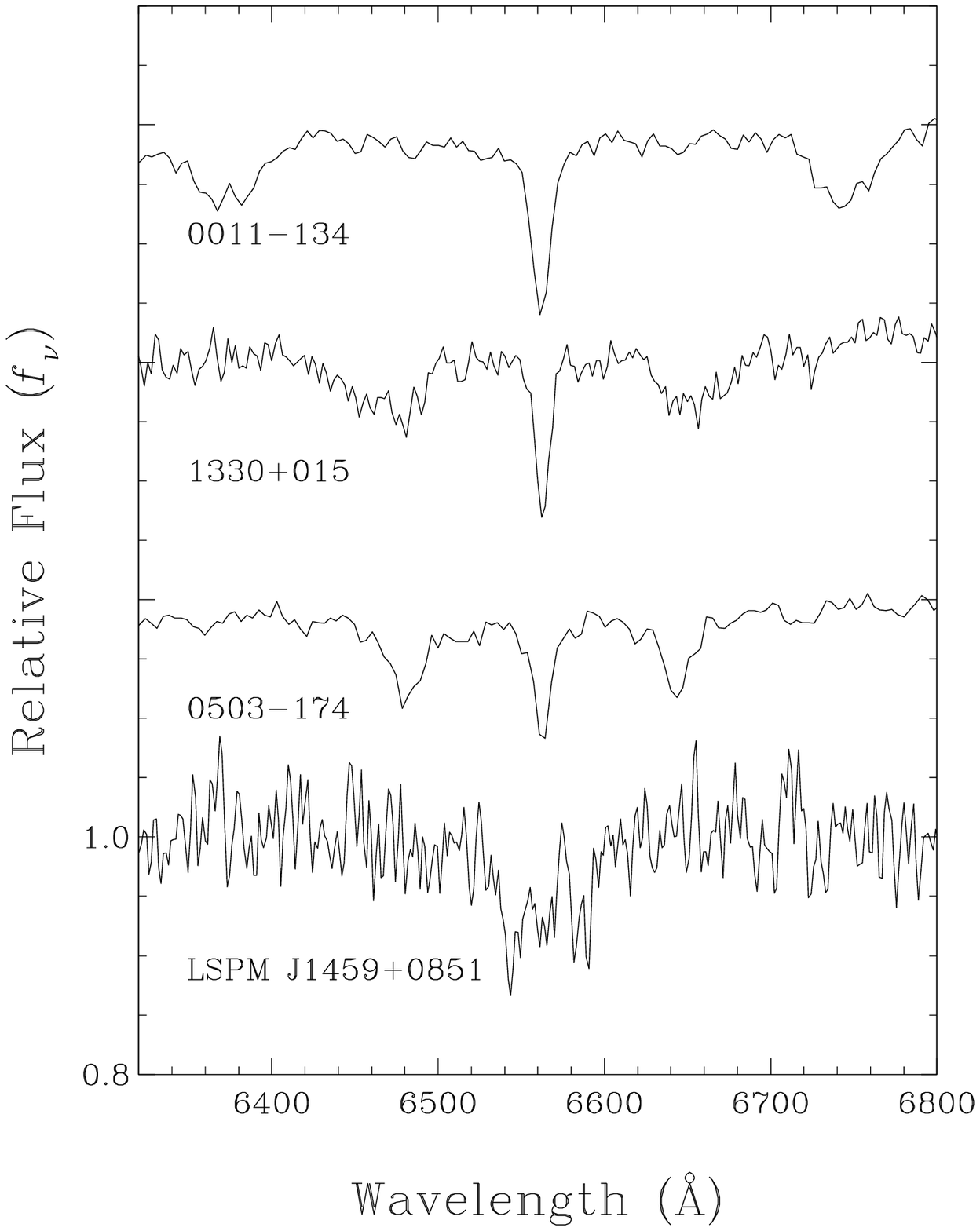}
\vspace{6.5cm}
\caption{Optical spectrum in the region of H$\alpha$ for the white dwarf LSPM J$1459+0851$. For comparison the spectra of three similar cool, hydrogen rich, magnetic white dwarfs, WD 0011-134 (B$\sim$16.7MG), WD 1330+015 (B$\sim$7.4MG)  and WD 0503-174 (B$\sim$7.3MG) are also shown, with magnetic field strength decreasing from top to bottom.}
\label{wdha}
\end{figure}

It can be seen that some residual tellurics remain and are highlighted for reference. They do not however affect the subsequent analysis in any way, since they do not overlap with features used directly to assess white dwarf properties. The general spectral shape is quite blackbody-like, consistent with a white dwarf or perhaps a very metal poor sub-dwarf (e.g. \citealt{jao08}). However the overall strength of the H$\alpha$ line and the peak of the blackbody-like continuum are only consistent with a relatively cool example of a white dwarf (\citealt{kilic06}). We also compare LSPM J$1459+0851$ to the spectra of three other very cool, hydrogen rich (DA) white dwarfs, WD0011-134, WD 1330+015 and WD 0503-174, taken from \citet{berg92} and \citet{berg93}. They have corresponding $T_{\rm eff}$'s of 6000$\pm$150\,K, 7450$\pm$200\,K and 5230$\pm$140\,K respectively, and are shown in Fig.~\ref{wdha}. Although the spectra of LSPM J$1459+0851$ are noisier, it can be seen that the extent of the H$\alpha$ feature is consistent with a cool, hydrogen rich white dwarf. We further assess its properties in more detail in the following sections.

\subsection{Synthetic Photometry \& Fitting Procedure}
\label{wdmodels}

As previously noted, the spectrum shows a lack of strong H lines,
which would be expected for a hotter more typical white dwarf. Photometric fitting
of the full optical-NIR SED is thus optimal for determining the effective temperature
of the white dwarf, using model fits to the SDSS+UKIDSS photometry and
assuming a distance equal to that of the T dwarf companion. The
spectra are consistent with a cool white dwarf, showing no strong
features in the spectrum blue-ward of 6000\AA\ and just a hint of
H$\alpha$. We performed a fit to the available photometry of the white dwarf
(from SDSS, USNO and UKIDSS) using the atmospheric model codes of
Bergeron et al., which are described at length in \citet[with updates
  given in \citealt{berg01,berg05}]{berg95}. These models assume local
thermodynamic equilibrium and allow energy transport by convection
and can be calculated with arbitrary amounts of hydrogen and
helium. Synthetic colors\footnote{The synthetic colors can be obtained
  at http://www.astro.umontreal.ca/$\sim$bergeron/CoolingModels/} were 
obtained using the procedure outlined in \citet{holberg06} based on
the Vega fluxes taken from \citet{bohlin04}.

The method used to fit the photometric data is similar to that
described in \citet{berg01}, which we briefly summarize here. We first
transform the magnitudes in each bandpass into observed average fluxes
$f^{m}_{\lambda}$ using the following equation

\begin{equation}
  m = $-$2.5 \log f^{m}_{\lambda} + c_{m}
\label{eq:1}
\end{equation}

\noindent where

\begin{equation}
  f^{m}_{\lambda} = \frac{\int^{\infty}_{0}f_{\lambda}S_{m}(\lambda)d\lambda}{\int^{\infty}_{0}S_{m}(\lambda)d\lambda}.
\label{eq:2}
\end{equation}

\noindent The transmission functions $S_{m}(\lambda)$ along with the
constants $c_{m}$ for each bandpass are described in
\citet{holberg06} and references therein. To make use of all the
photometric measurements simultaneously, we convert the magnitudes
into observed fluxes using equation (\ref{eq:1}) and compare the
resulting energy distributions with those predicted from our model
atmosphere calculations. Thus, we obtain a set of average fluxes
$f^{m}_{\lambda}$, which can now be compared with the model
fluxes. These model fluxes are also averaged over the filter
bandpasses by substituting $f_{\lambda}$ in equation (\ref{eq:2}) for
the monochromatic Eddington flux $H_{\lambda}$. The average observed
fluxes $f^{m}_{\lambda}$ and model fluxes $H^{m}_{\lambda}$, which
depend on $T_{\rm eff}$, $\log g$ and $N$(He)/$N$(H), are related by
the equation

\begin{equation}
  f^{m}_{\lambda} = 4\pi(R/D)^{2}H^{m}_{\lambda} 
\label{eq:3}
\end{equation}

\noindent where R/D is the ratio of the radius of the star to its
distance from Earth. Our fitting procedure relies on the nonlinear
least-squares method of Levenberg-Marquardt, which is based on a
steepest descent method. The value of $\chi^{2}$ is taken as the sum
over all bandpasses of the difference between both sides of equation
(\ref{eq:3}), properly weighted by the corresponding observational
uncertainties.  In our fitting procedure, we consider only $T_{\rm
  eff}$ and the solid angle free parameters.  As discussed in
\cite{berg01}, the energy distributions are not sensitive enough to
surface gravity to constrain the value of $\log g$, and thus for white
dwarfs with no parallax measurement, as is the case here, we simply
assume $\log g$ = 8.0, which is consistent with the distance estimate of the T dwarf companion. Our best fit for a pure-H atmosphere arises from a $T_{\rm eff}$ = 5535$\pm$45\,K, and is shown in Fig.~\ref{fg:f1}. 

\subsection{A magnetic spectrum}
\label{magwd}

The shape and weakness of the H$\alpha$ line gives a poor fit to a basic
 5535\,K model spectra, and we thus investigated the possibility that the
white dwarf could be magnetic with the H$\alpha$ line affected by Zeeman
splitting.  The line opacity was calculated as the sum of the
individual resonance-broadened Zeeman components. The line
displacements and strengths of the Zeeman components of H$\alpha$ are
taken from the tables of \citet{kemic74}, and the total line opacity
is normalised to that resulting from the zero-field solution. The
specific intensities at the surface, $I(\nu, \mu, \tau_{\nu}=0)$, are
obtained by solving the radiative transfer equation for various field
strengths and values of $\mu$ ($\mu=\cos\theta$, where $\theta$ is the
angle between the angle of propagation of light and the normal to the
surface of the star). In doing so, the polarization of the radiation
is neglected as we are mainly interested in the total monochromatic
intensity. The effect of the magnetic field on the continuum opacity
is also neglected. The emergent spectrum is then obtained from an
integration over the surface of the star ($H_{\nu} \propto \int
I_{\nu}\mu d\mu$) for a particular geometry of the magnetic field
distribution. We note that in this procedure, limb darkening is
explicitly taken into account because of the integration over $\mu$.

We use the offset dipole model to model the magnetic field of the
white dwarf \citep{achilleos89}. In this model, the magnetic field is
generated by a dipole. At the surface of the star (of radius unity),
the strength of the magnetic field is simply given by

\begin{equation}
  B = 0.5B_{d}(3\cos^{2}\theta + 1)^{1/2}
\label{eq:4}
\end{equation}

\noindent where $B_{d}$ is the dipole field strength, and $\theta$ is
the standard angle in spherical coordinates ($\theta$ = 0 at the
pole). For a given value of $B_{d}$, the flux received at the Earth
will also depend on the viewing angle $i$ between the dipole axis and
the line of sight ($i$ = 0 for a pole-on view). However, in this
particular model, the dipole is also offset from the center of the
star in an arbitrary direction. To simplify the calculation, we assume
the dipole is offset parallel to the dipole axis. In this case, the
value of the offset is measured from the center of the star and is
denoted $a_{z}$ (in units of stellar radius). Note that with the
offset dipole models, the value of the dipole field strength, $B_{d}$,
is no longer equal to the value of the polar field strength. 

\begin{figure}
\vspace{4.5cm}
\hspace{2.0cm}
\includegraphics{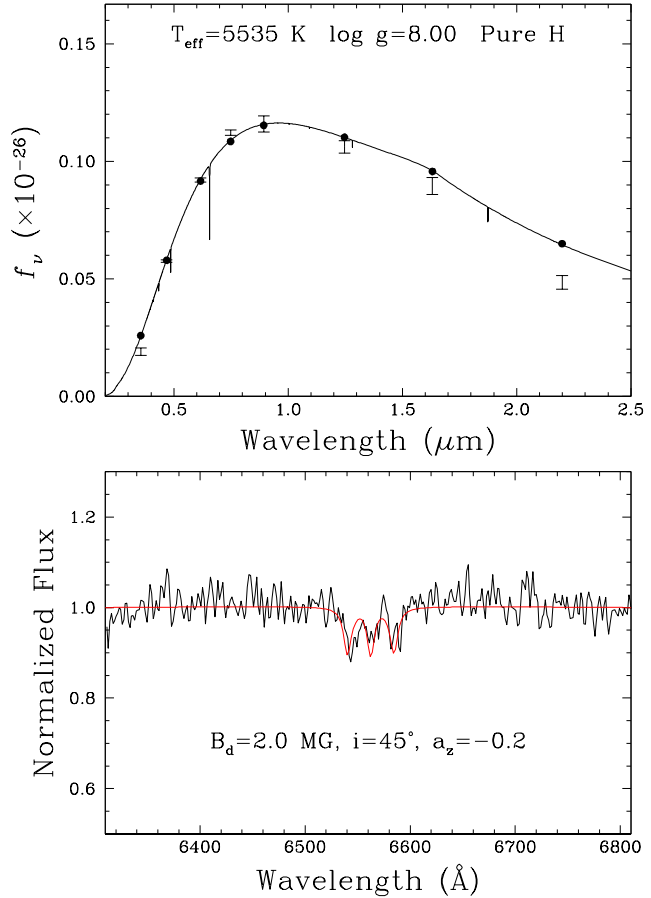}
\vspace{8.0cm}
\caption{{\it Top}: Fit of the energy distribution with pure-H models. The observed {\it ugriz} and {\it JHK} fluxes are represented by error bars while the model fluxes averaged over the filter bandpasses are indicated by filled circles. The model monochromatic fluxes are shown by a solid line. {\it Bottom}: Normalised spectrum near H$\alpha$ with the synthetic profiles interpolated at the parameters obtained from the energy distribution fits, assuming a pure-H atmospheric composition with a model fit at $T_{\rm eff}$ = 5535\,$\pm$45K and $\log g$ = 8.0 with an offset dipole magnetic field computed with the parameters indicated in the figure.}
\label{fg:f1}
\end{figure}

We computed a series of synthetic spectra based on a model atmosphere
of pure-H with $T_{\rm eff}$ = 5535\,K and $\log g$ = 8.0 while
varying $B_{d}$, $i$, and $a_{z}$. We display in Fig. \ref{fg:f1}
the model which best reproduced the observed line profile with $B_{d}$
= 2.0\,MG, $i$ = 45$^{\circ}$ and $a_{z}$ = $-$0.20. It should be noted
that varying the inclination angle $i$ produced only slight variations
in the line profiles and as stated in \citet{berg92}, it is not
possible to constrain $i$ from observed line profiles alone. The zeeman-split H$\alpha$ line in LSPM J$1459+0851$ can be compared to the same feature in other cool white dwarfs of higher magnetic field strengths in Fig~\ref{wdha}.

\subsection{White dwarf age and progenitor lifetime}
\label{wdage}

We start by considering the simple case of a non-magnetic white dwarf for a $T_{\rm eff}$ and $\log g$  of 5535\,K and 8.0, for which we calculate a mass of 0.585\,M$_\odot$  (\citealt{fontaine01}). The corresponding cooling age of a 0.585\,M$_\odot$ white dwarf, with a $T_{\rm eff}$= 5535\,K was then calculated as 3\,Gyr using the isochrones of \citet{fontaine01}. The total age of the white dwarf comprises of both the cooling time and its progenitor lifetime on the main-sequence. In order to constrain the main-sequence lifetime we accessed the Initial-final mass relations of \citet{ferrario05}, \citet{catalan08} and \citet{kalirai08} to estimate a likely, initial-mass constraint for the main-sequence progenitor star of 1.50-1.75\,M$_\odot$. We then used the tracks of \citet{Lach99} to estimate the main-sequence lifetime for stars of such mass as 1.8-3.0\,Gyrs. It should be noted however that the model tracks converge for masses $<$2\,M$_\odot$, and as a result the ages of these objects can be largely uncertain (up to 10\,Gyr).

The strong magnetic field present also provides an additional factor to consider when assessing age. The origin of such strong white dwarf magnetic fields is not fully understood but is thought to have arisen in one of two favored scenarios: 

{\it 1.} From a single star. The magnetic field is thought to derive from a massive, magnetic progenitor of $\sim$1.5-8\,M$_\odot$, typically an Ap or Bp star. The magnetic field is then maintained through the main-sequence evolution to the white dwarf phase by flux conservation (\citealt{wick00}).

{\it 2.} From the merger of two stellar cores in a common envelope event or the merger of two degenerate objects. During the common envelope (CE) phase the orbits of the two cores spiral in closer together through frictional forces causing differential rotation, which coupled with convection in the cores, creates a stellar-magnetic dynamo (\citealt{tout92}). Close, but separated cores form CVs but some cores coalesce and cool to form a magnetic white dwarf. It may also be possible that two very close white dwarfs emerge from the CE phase, such as G62-46 (\citealt{berg93}), where one component is highly magnetic.

In general it is observed that magnetic white dwarfs have larger masses than the more typical non-magnetic white dwarfs (e.g. \citealt{liebert88}). There are two possible hypothesis to explain this. Firstly, in accordance with the favoured scenario for the formation of isolated magnetic white dwarfs (\citealt{wick05}), the progenitor was more massive, leading to a massive white dwarf. In this case the magnetic field has no effect during the progenitor evolution. Secondly the effect of the magnetic field has an impact on the stellar evolution, such that it could inhibit mass loss (\citealt{wick00}), leading to a more massive core and a longer progenitor lifetime. If we consider the possibility that our white dwarf could be of higher mass, for example 0.8M\,$_\odot$ (the mean mass of a highly magnetic white dwarf; \citealt{kawka07}) then the cooling time would be longer, around 6\,Gyrs (\citealt{fontaine01}). In this case the progenitor (if a single star) would be around 3.5\,M$_\odot$ (\citealt{catalan08}) and the progenitor main-sequence lifetime would be 0.3\,Gyr (assuming normal models). However there is also evidence that many magnetic white dwarfs have masses closer to the peak of the non-magnetic mass distribution (\citealt{tout08}).
 
As we cannot know which scenario is responsible for the observed magnetic field, nor can we measure more accurately the mass of the white dwarf, we do not know the effects this may have had on the main-sequence evolution. In any case both scenarios for a magnetic and a non-magnetic white dwarf results in ages greater than 4.8\,Gyrs (the cooling age plus the minimum progenitor lifetime of a non-magnetic white dwarf) and we thus choose to adopt this as the minimum total age for LSPM J$1459+0851$.

\begin{table}
\caption{Parameters of the white dwarf LSPM J$1459+0851$.} \centering
\begin{tabular}{|l|l|c|}
\hline
Parameter &  & Value \\
\hline
RA & ............ & 14 59 32.05\\
DEC & ............ & +08 51 28.1\\
SDSS $'u$  & ............ & 20.74\,$\pm$ 0.08\\
SDSS $'g$  & ............ & 19.50\,$\pm$ 0.01\\
SDSS $'r$  & ............ & 18.99\,$\pm$ 0.01\\
SDSS $'i$  & ............ & 18.76\,$\pm$ 0.05\\
SDSS $'z$  & ............ & 18.71\,$\pm$ 0.03\\
SuperCOSMOS $B$  & ............ & $\sim$19.48\\
SuperCOSMOS $R$  & ............ & $\sim$18.33\\
SuperCOSMOS $I$  & ............ & $\sim$18.29\\
USNO $B$  & ............ & $\sim$19.8\\
USNO $R$  & ............ & $\sim$18.8\\
USNO $I$  & ............ & $\sim$18.2\\
UKIDSS $Y$  & ............ & 18.14\,$\pm$ 0.02\\
UKIDSS $J$  & ............ & 17.90\,$\pm$ 0.02\\
UKIDSS $H$  & ............ & 17.65\,$\pm$ 0.04\\
UKIDSS $K$  & ............ & 17.79\,$\pm$ 0.06\\
$\mu$ RA  & ............ & $-$170\,$\pm$3\,mas yr$^{-1}$$^*$\\
$\mu$ DEC  & ............ & $-$42\,$\pm$6\,mas yr$^{-1}$$^*$\\
$T_{\rm eff}$  & ............ & 5535$\pm$45\,K\\
$\log g$ & ............ & $8.0$\,dex\\
Mass  & ............ & $0.585$\,M$_{\odot}$\\
WD age & ............ & $>$4.8\,Gyr\\
\hline
\multicolumn{3}{l}{$^*$ USNO-B1}\\
\hline
\end{tabular}
\label{properties}
\end{table}

\section{LSPM J1459+0851 - ULAS\,J1459+0857 a bound system?}
\label{stats}

In order to determine if this new system is a bonafide binary system,
we have statistically assessed the likelihood that two such objects
could be a line-of-sight association with photometry and proper motion
consistent with binarity by random chance. We first calculated the
total region of sky around our T dwarf corresponding to the coverage 
encompassed by the projected line-of-sight separation of the white dwarf. We then combined 
this with the T dwarf distance constraint (43-69pc), allowing for the 
possibility that the T dwarf might itself be an unresolved binary (at 
a greater distance), to estimate a volume of sky in which white dwarfs 
might be line-of-sight contamination. We then used the number density 
of white dwarfs (e.g. \citealt{schroeder04}) to estimate 
that we would expect only 0.00368 white dwarfs in this volume of space.

To factor in the probability that two objects might have a common
proper motion at the level of our measurements, we downloaded a
magnitude-limited sample ($R$\,$<$21, the same as our initial selection; see Section \ref{cpm}) of objects from the SuperCOSMOS Science
Archive. We applied a limit to the proper motion uncertainty of 
$<$50mas/yr and required objects to lie in the colour range 1$<B-R<$3 (where we expect contaminant, main-sequence stars). This sample 
of 10,360 sources was selected from within one degree of the T dwarf as to provide a representative sample of objects in the area of sky in which we find our binary system. We then placed these objects on a colour-magnitude diagram and selected only objects that occupied a region populated by main-sequence stars, when placed at 
the distance range estimated for the T dwarf. Of the 140 objects that were
selected in this way, 13 had proper motion consistent (at the 1\,$\sigma$ level) 
with the T dwarf, leading to a probability of 0.092, that a contaminant 
star could have proper motion common with the T dwarf.

\begin{figure*}
\includegraphics[width=150mm, angle=0]{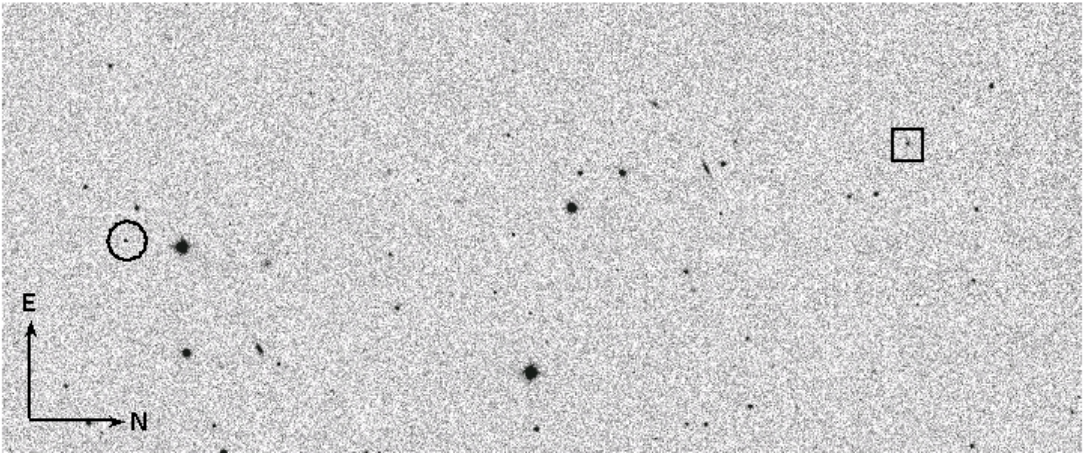}
\caption{A UKIDSS $J$ band finder chart showing the position of the T dwarf (square) and the white dwarf (circle), the scale of the image is 8$'$ x 3.5$'$. }
\label{finder}
\end{figure*}

The chance of finding a white dwarf with the same line-of-sight separation 
of LSPM\,J1459+851 from ULAS\,J1459+0857, where both components share a common proper motion and are consistent with lying at the same distance
as that estimated for the T dwarf is thus 0.00368 x 0.092 = 0.0003. We thus conclude that these objects form a genuine binary system. Since we searched for companions to a total of 19 T dwarfs, we estimate the overall chance of finding a spurious system in our sample is 0.0064, and that the systems identified are likely real binaries. Properties of the new white dwarf + T dwarf 
system are given in Table.~\ref{binproperties}, and a finding chart 
is presented in Fig.~\ref{finder}. The wide separation of the system (16,500AU, assuming it is a singular object and not an unresolved binary)
is similar to the widest known BD+main sequence binary systems 
(e.g. \citealt{faherty09}, \citealt{zhang10}). Although prior to the 
post-main-sequence mass loss phase of the primary the separation would have been substantially less. Indeed we expect that the initial separation of the two components was a factor of $\sim$4 closer (see Section \ref{cpm}), in the region of$\sim$4100AU and more akin to the more common 
type of BD+main sequence binaries ($<$5000AU; \citealt{pinfield08}).

\begin{table}
\caption{Parameters of the binary system.} \centering
\begin{tabular}{|l|l|c|}
\hline
Parameter &  & Value \\
\hline
Separation on sky & ............ & 385 arcsec\\
Estimated distance & ............ & $43-69$\,pc$^{*}$\\
Estimated line-of-sight \\
separation & ............ & $16,500-26,500$\,AU$^{*}$\\
\hline
\multicolumn{3}{l}{$^{*}$ Assuming the T dwarf is a singular or unresolved binary. }\\
\hline
\end{tabular}
\label{binproperties}
\end{table}

\section{Properties of ULAS\,J1459+0857}
\label{tproperties}

In order to estimate the $T_{\rm eff}$ of ULAS\,J1459+0857 we used 
spectral type-$T_{\rm eff}$ determinations from Table 6 of \citet{golimowski04} as a guide. These determinations are for BDs with known parallaxes (e.g. \citealt{tinney96}; \citealt{leggett02}; \citealt{dahn02}; \citealt{tinney03}; \citealt{knapp04}; \citealt{vrba04}) and SED constraints 
over a broad wavelength range, for which reliable bolometric flux 
measurements and luminosities are thus available. The main source 
of uncertainty in these $T_{\rm eff}$ values comes from a lack of strong 
age constraints and the resulting evolutionary model radii uncertainties 
(up to $\sim$30\% for ages $>$0.1 Gyrs; \citealt{burrows97}; \citealt{baraffe98}; \citealt{chabrier00}). By considering the variety of $T_{\rm eff}$ 
ranges calculated for the T4.5$\pm$0.5 dwarfs which have assumed a range of possible ages from 0.1-10\,Gyrs ($\pm\sim$300K), we estimate that for an age range of 4-10\,Gyrs ULAS\,J1459+0857 has $T_{\rm eff}$ in the range 1200-1500K.

\begin{figure*}
\includegraphics[width=100mm, angle=90]{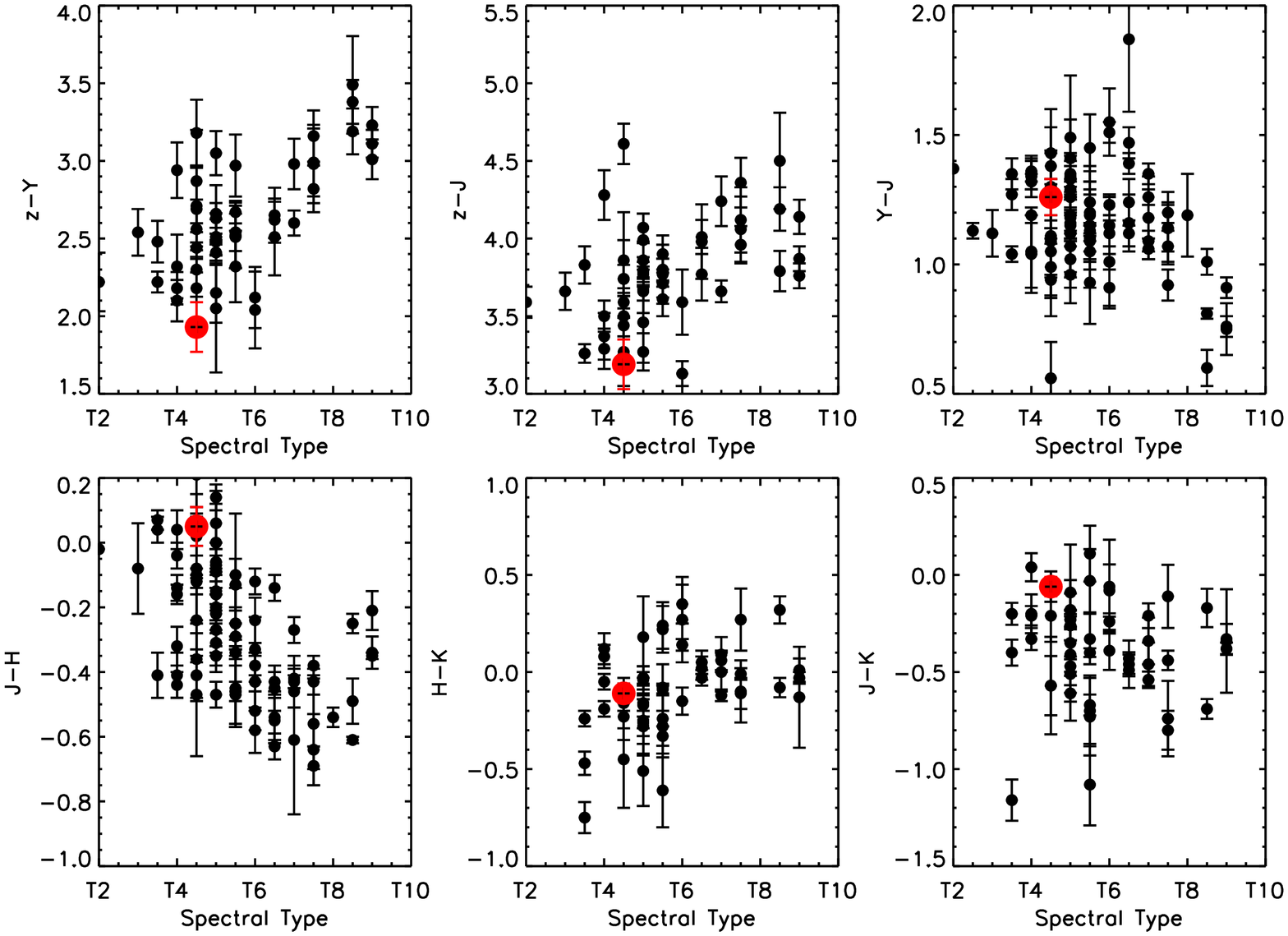}
\caption{Colour against T dwarf spectral type for UKIDSS T dwarfs (\citealt{burningham10}). ULAS\,J1459+0857 is shown as a large, filled red circle. }
\label{tcols}
\end{figure*}

We also used the Lyon Group COND models \citep{baraffe03} to estimate
the physical properties of ULAS\,J1459+0857, allowing for the
possibility that it could be a single object or itself an unresolved
binary, with a distance in the range 43-69pc. We calculated absolute magnitudes for ULAS\,J1459+0857 based on this distance range for ages 4-10\,Gyrs. Then using a linear interpolation between the model grid points we obtained mass and $\log{g}$ estimates in the range 0.06-0.072M$_\odot$ and 5.4-5.5\,dex respectively,
assuming solar metallicity. We also consider the evolutionary models
of Burrows et al. (\citealt{burrows97}; \citealt{burrows01}; \citealt{burrows06}) to
estimate a mass of 0.064-0.075M$_\odot$ if the T dwarf is actually metal poor
([M/H]~-0.5\,dex), which is a similar value to that of the solar metallicity COND models. Both models also indicate a high gravity ($\log{g} =$ 5.5) for the observable $J-K$ colours and temperature range of ULAS\,J1459+0857. We also compare the optical + NIR colours of ULAS\,J1459+0857 in comparison with other spectroscopically confirmed T dwarfs from the UKIDSS LAS \citep{burningham10}. Fig.~\ref{tcols} shows this compliment of T dwarfs on a series of colour-spectral type plots. Whilst the NIR colours of ULAS\,J1459+085 in general look fairly typical for a T4.5 dwarf, there is some evidence of relative $z-$ band enhancement, which could be the result of the older, higher gravity nature of this object \citep{pinfield08}.

\subsection{Initial model testing with benchmark observations}
\begin{figure*}
\includegraphics[width=100mm, angle=-90]{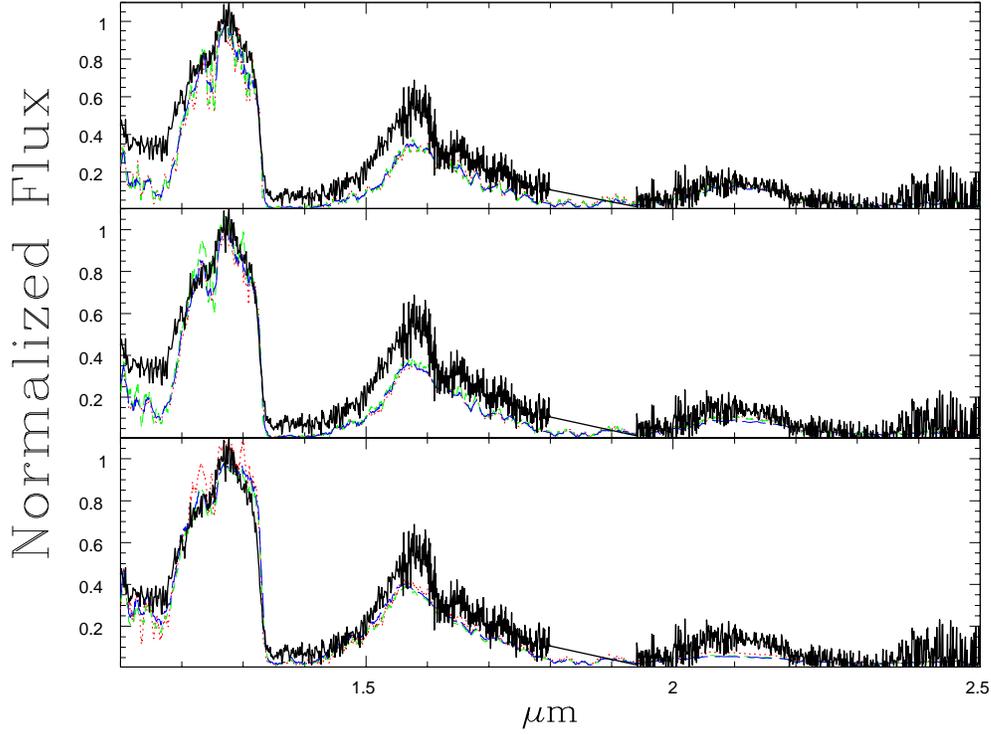}
\caption{Spectral model comparisons to ULAS\,J1459+0857 (black solid line) for  $T_{\rm eff}$=1200\,K and [M/H]= +0.3, 0.0 and $-$0.5, from top to bottom, with $\log{g}$ = 5.0, 5.25 and 5.5 as a long red dotted line, short dashed green and the long dashed blue lines, respectively. }
\label{tdspec1200}
\end{figure*}

\begin{figure*}
\includegraphics[width=100mm, angle=-90]{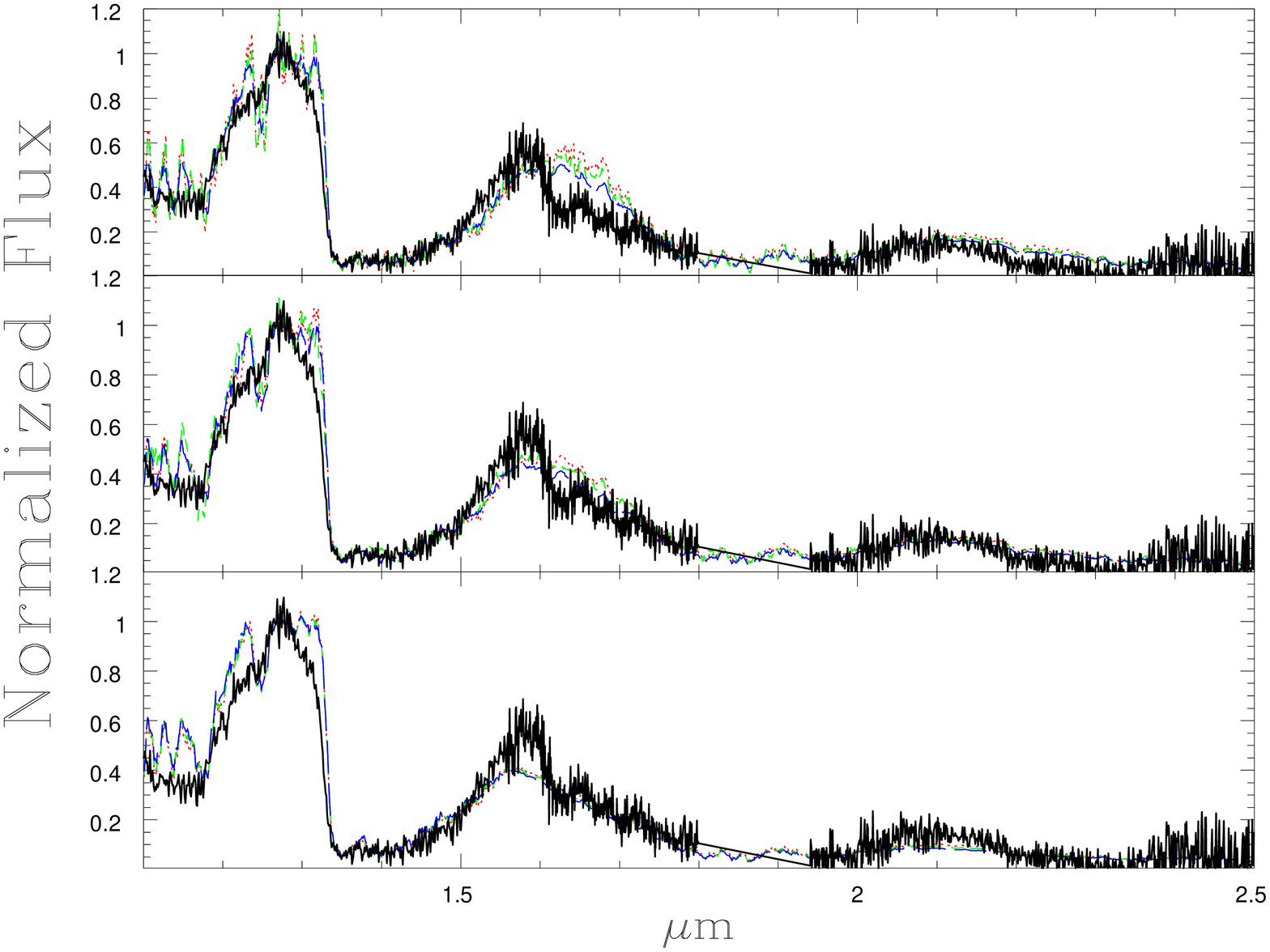}
\caption{Spectral model comparisons to ULAS\,J1459+0857 (black solid line) for  $T_{\rm eff}$=1500\,K and [M/H]= +0.3, 0.0 and $-$0.5, from top to bottom, with $\log{g}$ = 5.0, 5.25 and 5.5 as a long red dotted line, short dashed green and the long dashed blue lines, respectively.}
\label{tdspec1500}
\end{figure*}

To provide a first-pass test of model atmosphere predictions, we
used the dust-free COND models of \citet{baraffe03} to provide theoretically
informed best-guess constraints of the physical properties of the
T dwarf in our binary. We made comparisons between the observed T dwarf
spectrum and synthesised spectroscopy for two values of $T_{\rm eff}$
(1200 and 1500\,K), three values of $\log g$ (5.0, 5.25 and 5.5) and
three metallicities (+0.3, 0.0 and $-$0.5, representing the range observed
in the galactic disk;  e.g. \citealt{valenti05};
\citealt{gray06}; \citealt{holmberg07}; \citealt{jenkins08}). The model spectra were generated using the
atmospheric radiative transfer code, Phoenix (which is described in detail in \citealt{haus99};
\citealt{allard01}) which includes the updated water molecular
opacity list from \citet{barber07} and new solar abundances from
\citet{asplund09}. The model includes the effect of condensation in
the chemical equilibrium but ignores the effects of dust opacities.
Theoretical spectra were normalised to the observations in the peak
of the $J-$ band, and the resulting comparisons are shown in
Fig.~\ref{tdspec1200} and \ref{tdspec1500}. The long red dotted line, short dashed green
and the long dashed blue lines represent the $\log{g} =$ 5.0, 5.25
and 5.5, respectively.

Table.~\ref{chifit}  gives a summary of the reduced chi-squared ($\chi$$^2_{\nu}$)  values calculated
for these comparisons. These values range from $\sim$1.75-3.5, and we note
that none of them are close to 1.0. This is to be expected since there
are known shortcomings in the models that introduce differences
significantly greater than the measurement uncertainties - e.g. incomplete methane
opacities and a poor understanding of the observed $J-$band brightnening
around T3 (e.g. \citealt{knapp04}). As such, we only use these $\chi$$^2_{\nu}$ values as an additional
quality indicator for the over-all model-observation comparison. 

Our $\chi$$^2_{\nu}$ analysis reveals that the best fitting atmospheric model has an effective temperature of 
1200K, a sub-solar metallicity and a high gravity of $\log{g} =$5.5.  However, comparison by eye shows clearly that 
this fit across the whole spectral range is not particularly good, especially around the peak flux regions of the 
$H-$ and $K-$ bands. This best $\chi$$^2_{\nu}$ value presumably comes from the better fit to the $J-$ band, compared to the other models, however a better fit (by eye)  to the overall profile of the spectra comes from a 1500K, solar metallicity, high gravity model. This suggests that  ULAS\,J1459+0857 probably has solar to slightly sub-solar metallicity and high gravity. This is instructive at least that the observed properties of ULAS\,J1459+0857 appear to be in general agreement with those predicted by evolutionary models. While the models clearly have some shortcomings they appear to be making progress, such that model fit properties for a mid T dwarf of age $>$ 4\,Gyrs seem to be broadly consistent with the benchmark fit properties.

\begin{table}
\caption{ $\chi$$^2_{\nu}$ fits to model spectra.}
\begin{tabular}{|l|c|c|c|}
\hline
$\log{g}$ &  & $[M/H]$ & \\
          & +0.3 & 0.0 & $-$0.5 \\
\hline
$T_{\rm eff}$=1200\,K\\
\hline
5.00 &       1.94 & 2.11 & 2.38\\
5.25 &       2.06 & 2.58 & 1.82\\
5.50 &       1.97 & 2.07 & 1.76\\
\hline
$T_{\rm eff}$=1500\,K\\
\hline
5.00 &       3.50 & 2.47 & 2.55\\
5.25 &       2.89 & 2.03 & 2.62\\
5.50 &       3.09 & 1.96 & 2.88\\
\hline
\end{tabular}
\label{chifit}
\end{table}

\begin{table}
\caption{Parameters of ULAS J$1459+0857$.} 
\centering
\begin{tabular}{|l|l|c|}
\hline
Parameter &  & Value \\
\hline
RA & ............ & 14 59 35.25\\
DEC & ............ & +08 57 51.20\\
Distance & ............ & 43 -- 69pc$^*1$\\
SDSS $z'$  & ............ & 21.17\,$\pm$ 0.15\\
UKIDSS $Y$  & ............ & 19.24\,$\pm$ 0.06\\
UKIDSS $J$  & ............ & 17.93\,$\pm$ 0.02\\
UKIDSS $H$  & ............ & 17.94\,$\pm$ 0.05\\
UKIDSS $K$  & ............ & 17.92\,$\pm$ 0.08\\
UFTI $J$ & ............. &17.98\,$\pm$ 0.04\\
UFTI $H$ & ............. &17.93\,$\pm$ 0.04\\
UFTI $K$ & ............. &18.04\,$\pm$ 0.03\\
$\mu$ RA  & ............ & $-$149\,$\pm$33\,mas yr$^{-1}$\\
$\mu$ DEC  & ............ & $-$45\,$\pm$33\,mas yr$^{-1}$\\
Spectral Type  & ............ & T4.5\,$\pm$0.5\\
Mass & ............ & $0.060-0.072$\,M$_\odot$$^{*2}$\\
Mass & ............ & $0.064-0.075$\,M$_\odot$$^{*3}$\\
$T_{\rm eff}$ & ............ & $1200-1500$\,K\\
$\log{g}$ & ............  & $5.4-5.5$\,dex \\
\hline
\multicolumn{3}{l}{$^{*1}$ Assuming the T dwarf is a singular or unresolved binary. }\\
\multicolumn{3}{l}{$^{*2}$ From the Lyon group COND evolutionary models. }\\
\multicolumn{3}{l}{$^{*3}$ From the Burrows evolutionary models. }\\
\hline
\end{tabular}
\label{propertiest}
\end{table}

\section{Summary}
\label{concs} 

This is the first discovery of a T dwarf + white dwarf binary
system, and an example of an evolved, high gravity brown dwarf. 
During the main sequence phase of the primary its separation would 
have been similar to the bulk population of BD+main sequence 
binaries, although the separation must have grown significantly 
(to its current value) during the post-main-sequnce mass loss 
phase of the primary.

The white dwarf provides vital age constraints for the binary system 
from it's cooling age combined with a minimum estimate of it's main 
sequence progenitor lifetime, and the resulting minimum age of 
4.8\,Gyrs for the T dwarf allows a robust constraint on its surface 
gravity of $\log{g}=5.4-5.5$. As such ULAS\,J1459+0857 is a representative 
old, high gravity benchmark BD. Comparison with the bulk 
popualtion of UKIDSS T dwarfs shows some indication that $z-$ band flux enhancement may 
be an observational characteristic of older high gravity mid T dwarfs, 
which could be a useful factor in attempts to understand the formation 
history of sub-stellar objects. And more generally, this T dwarf can 
contribute to our understanding of substellar properties by providing 
a useful test-bed for theoretical model atmospheres.

This system is an 
example of how wide BD binary companions to white dwarfs make good
benchmark objects, which will help test model atmospheres, and may
provide independent means to calibrate BD properties of field objects
(\citealt{pinfield06}). This is the first of our candidate systems and
we expect to find many more as we mine more UKIDSS sky, and with our
ongoing efforts to search still deeper by combining UKIDSS BDs with
SDSS for white dwarfs.

Further observations that would have clear benefits for this system include 
a parallax measurement of either component to yield an accurate distance, 
and fuller spectral coverage of the T dwarf to constrain its mid-infrared 
and optical spectral morphology.  In general, and particularly in the optical $z-$band, it would be desirable to assess in more detail spectroscopic features and trends that may be sensitive to higher surface gravity and age.  A parallax distance combined with accurate  knowledge of the T dwarf bolometric flux would offer significant improvements  on the $T_{\rm eff}$ constraints for this object (e.g. \citealt{burningham09}). 
 
An accurate distance would also facilitate greatly improved radius and 
mass constraints for the white dwarf, and thus a better constraint on 
the cooling age and progenitor lifetime. More detailed studies of both 
binary components are clearly crucial to maximize the effectiveness of 
the benchmark BD component.

\section*{Acknowledgments}
ADJ, JJ and MTR would like to acknowledge the support of the grant from
CONICYT and the partial support from Center for Astrophysics FONDAP
and Proyecto Basal PB06 (CATA). This work was also suported in part by
the NSERC Canada and by the Fund FQRNT (Qu\'ebec). P.B. is a Cottrell
Scholar of Research Corporation for Science Advancement. JG is
supported by RoPACS, a Marie Curie Initial Training Network funded by
the European Commission's Seventh Framework Programme. SC is supported by a Marie Curie Intra-European Fellowship within the 7th European
Community Framework Programme.

This publication has made use of the the data obtained from the
SuperCOSMOS Science Archive, prepared and hosted by the Wide Field
Astronomy Unit, Institute for Astronomy, University of Edinburgh,
which is funded by the STFC. We also acknowlegde the use of data from
the Sloan digital archive, which is funded by the Alfred P. Sloan
Foundation, the Participating Institutions, the National Aeronautics
and Space Administration, the National Science Foundation, the
U.S. Department of Energy, the Japanese Monbukagakusho, and the Max
Planck Society. We have used data from the Large Area Survey, including those from the
data release 4, \citep{warren07}. We also acknowledge the use of DENIS and the SIMBAD database, operated at CDS, Strasbourg, France.

\bsp

\label{lastpage}

\end{document}